\documentclass[prl,twocolumn,english,superscriptaddress,floatfix]{revtex4}

\usepackage{graphicx}
\usepackage{float}
\usepackage{bm, amsmath, amssymb}
\usepackage{color}
\usepackage{soul}
\usepackage{pdfpages}
\usepackage[T1]{fontenc}

\usepackage[utf8]{inputenc}
\usepackage{braket}
\usepackage{xfrac}
\usepackage{verbatim}
\usepackage{lipsum}
\usepackage{cancel}

\usepackage{amstext}
\usepackage{bbold}
\usepackage{esint}
\usepackage{babel}
\usepackage{relsize}
\newcommand{\addkm}[1]{\textcolor{black}{#1}}

\newcommand{\revkm}[1]{\textcolor{black}{#1}}

\newcommand{\be}{\begin{equation}}
\newcommand{\ee}{\end{equation}}
\newcommand{\ra}{\rangle}
\newcommand{\la}{\langle}

\begin{document}

\title{Achieving optimal quantum acceleration of frequency estimation using adaptive coherent control}

\author{M. Naghiloo}
\affiliation{Department of Physics, Washington University, St.\ Louis, Missouri 63130}
\author{A. N. Jordan}
\affiliation{Department of Physics and Astronomy, University of Rochester, Rochester, NY 14627}
\affiliation{Center for Coherence and Quantum Optics, University of Rochester, Rochester, NY 14627}
\affiliation{Institute for Quantum Studies, Chapman University, Orange, CA 92866, USA}
\author{K. W. Murch}
\affiliation{Department of Physics, Washington University, St.\ Louis, Missouri 63130}
\affiliation{Institute for Materials Science and Engineering, St.\ Louis, Missouri 63130}

\date{\today}

\begin{abstract}
Precision measurements of frequency are critical to accurate timekeeping, and are fundamentally limited by quantum measurement uncertainties.  While for time-independent quantum Hamiltonians, the uncertainty of any parameter scales at best as $1/T$, where $T$ is the duration of the experiment, recent theoretical works have predicted that explicitly time-dependent Hamiltonians can yield a $1/T^2$ scaling of the uncertainty for an oscillation frequency.  This quantum acceleration in precision requires coherent control, which is generally adaptive. We experimentally realize this quantum improvement in frequency sensitivity with superconducting circuits, using a single transmon qubit.  With optimal control pulses, the theoretically ideal frequency precision scaling is reached for times shorter than the decoherence time.  This result demonstrates a fundamental quantum advantage for frequency estimation.
\end{abstract}

\maketitle

The ability to sense more accurately has historically been the basis of many of our scientific advances and technological innovations.  In particular, precision measurements have been instrumental in advancing our knowledge of fundamental physical laws \cite{Abe2008, Walk1994, Agui2013, rosi2014}. Notably, frequency measurements have been essential to experimental tests of general relativity, the standard model of particle physics, and quantum mechanics, and are the practical foundation of all timekeeping devices.   The precision of measurements is ultimately governed by the fundamentally probabilistic nature of quantum measurements, which arises most basically in the Heisenberg uncertainty principle.  Traditionally, frequency measurements, such as are conducted with atomic clocks \cite{essen1955atomic,lyons1957atomic,ludl15rmp}, are associated with the measurement of the energy difference, $E$, between two eigenvalues of a static Hamiltonian $H$, and the frequency uncertainty arises from the energy-time uncertainty principle \cite{gabor1946theory}, $\delta \omega = \delta E/\hbar   = 1/(2T)$, where $T$ is the time of the experiment. New situations arise in frequency metrology when one considers instead  \emph{time-dependent} Hamiltonians, where the precision of frequency measurements can be optimized with additional control. 

   
In metrology, one seeks to determine a parameter $g$ from repeated measurements that naturally follow a probability distribution $p_g(X)$, where $X$ is some random variable.  For large data sets, the Cram\'er-Rao bound \cite{cram46} gives a universal limit for the mean squared deviation of the parameter, 
\begin{align}
\langle \delta^2 {\hat g} \rangle \geq \frac{1}{v I_g}, 
\end{align}
where $v$ is a measure of the amount of data, ${\hat g}$ is an unbiased estimator of the parameter $g$ formed from measurement data, and $I_g = \int p_g(X) (\partial_g \ln p_g(X))^2 dX$ is the Fisher information \cite{fish25} which characterizes the amount of information about the parameter $g$ that is contained in the data.  Therefore, the Fisher information is a natural measure of how optimal a given measurement strategy is for determining the parameter $g$ with minimal uncertainty.  

For quantum parameter estimation, measurements on quantum states $|\psi_g\rangle$ are used to find the probability distribution $p_g(X)$. In this case  the Fisher information in the quantum state is given by \cite{Brau94,Brau96},
\begin{align}
I_g^{(Q)} = 4 \left( \langle \partial_g \psi_g | \partial_g \psi_g \rangle- |\langle \psi_g | \partial_g \psi_g \rangle |^2 \right), \label{eq:qfi}
\end{align}
which maximizes the classical Fisher information in the measurement results on the state over all possible types of quantum measurements.  The quantum Fisher information is a measure of the distinguishability of two states $|\psi_g \rangle $ and $|\psi_{g+ \mathrm{d}g} \rangle$ and with this formulation, it is clear that some quantum states garner more quantum Fisher information than others. In particular non-classical correlations can enhance measurement sensitivities.   The use of such non-classical resources in measurement has been widely studied \cite{Alip2014, Tsan2011, giov04} and applied in several metrological areas including imaging \cite{kolo99}, gravitational waves \cite{ligo11}, and magnetometry \cite{sewe12}.  Much of this research has focused on the scaling of the quantum Fisher information with the number $N$ of quantum systems; whereas uncorrelated systems lead to the standard quantum limit $\propto N$, appropriate quantum correlations can lead to the Heisenberg scaling $\propto N^2$ \cite{giov04}.

\begin{figure*}
\begin{center}
\includegraphics[width = 0.7\textwidth]{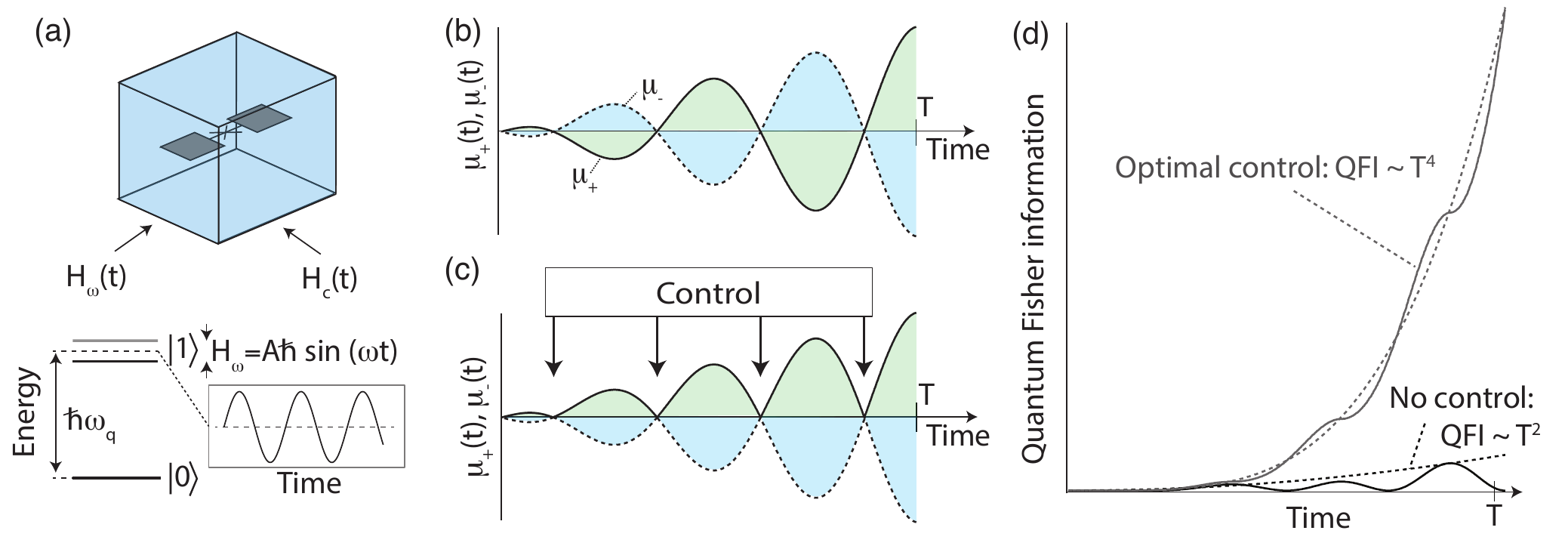}
\end{center}
\caption{\small {\bf Frequency estimation of a time-periodic Hamiltonian.} (a), The experiment consists of a transmon qubit dispersively coupled to a waveguide cavity. The qubit is subject to a time-dependent Hamiltonian $H_\omega(t)$ and the task is to estimate the frequency $\omega$.  (b), The eigenvalues $\mu_\pm$ of \revkm{$\partial_\omega H_\omega(t)/\hbar$}. The quantum Fisher information is related to the integral of $\mu_+(t) - \mu_-(t)$, which is alternately positive or negative.  (c), A control $H_\mathrm{c}(t)$ is used to guide the qubit evolution such that $\mu_+(t)$ and $\mu_-(t)$ are maximally separated. (d), The scaling of the quantum Fisher information for the uncontrolled and controlled measurement evolution, showing scaling as $T^2$ and $T^4$ respectively.   }
\label{fig1}
\end{figure*}

Here, we focus rather on how the quantum Fisher information for a single quantum system scales with time \cite{itan93}. If, for example, the parameter to be estimated is a multiplicative factor \cite{giov06} on a static Hamiltonian, $H_g = g H_0$, then, given that unitary evolution for a time $T$ is described by $U_g = \exp(-igH_0 T)$, the quantum Fisher information scales in time as $I_g \propto T^2$ \cite{giov06}.  However, if the Hamiltonian is instead time-dependent \cite{Tsan2011, cler15} the quantum Fisher information may exceed this scaling for certain parameters, reaching a scaling of $I_g \propto T^4$ for estimating the frequency of an oscillating Hamiltonian under optimal coherent control \cite{pang17}. \addkm{Very recent experiments \cite{boss17, schm17, jord17} in magnetic field sensing NV centers have demonstrated that using a hybrid quantum/classical strategy of estimating a local magnetic field value with a quantum technique, repeated in time, together with a classical Fourier transform can achieve a Fisher information of the frequency scaling as $T^3$.}
In this \addkm{letter}, we experimentally demonstrate $T^4$ scaling of the quantum Fisher information in the estimation of a Hamiltonian oscillation frequency for a pseudo-spin half system. This quantum enhanced scaling has been proved \cite{pang17} to be the best allowed by quantum mechanics in the kind of system we consider in this work. This improved scaling is achieved through adaptive optimal control where an additional control Hamiltonian that depends on the estimated parameter is applied to the system to enhance sensitivity.   We show that the $T^4$ scaling is robust against small variations in the control Hamiltonian, thus allowing for adaptive control.

To illustrate how optimal control can be used to maximize the quantum Fisher information, we consider a time-dependent Hamiltonian imposed on a two-level quantum system $H_\omega(t) = A \hbar \sin(\omega t)\sigma_z/2$, describing periodic modulation of the energy levels of the system with amplitude $A$ as shown in Figure 1a. Our focus is to maximize the quantum Fisher information of the modulation frequency $\omega$, that is to minimize the overlap of two quantum states $|\psi_\omega \rangle$ and $|\psi_{\omega+\delta \omega} \rangle$ after time evolution under the Hamiltonian for time $T$.   We will show that the optimal choice of quantum states is a superposition of energy eigenstates $(|0\rangle + e^{i\phi} |1\rangle)/\sqrt{2}$, which accumulate different phases $\phi_\omega(T)$ under the Hamiltonian evolution.

To formalize our discussion of the quantum Fisher information, we reformulate Equation~\ref{eq:qfi} as $I_\omega^{(Q)} = 4 \mathrm{Var}[h_\omega(T)]_{|\psi_0\rangle}$, where $h_\omega(T) = i U_\omega^\dagger(0\rightarrow T) \partial_\omega U_\omega(0\rightarrow T)$, $U_\omega(0\rightarrow T)$ is the unitary evolution of the initial state $|\psi_0\rangle$ under the Hamiltonian, and $\mathrm{Var}[\cdot]$ represents the variance.  In this form, we can see that the  quantum Fisher information is related to the squared difference between the minimum and maximum eigenvalues of $h_\omega(T)$. 

 \begin{figure*}
\begin{center}
\includegraphics[width = .9\textwidth]{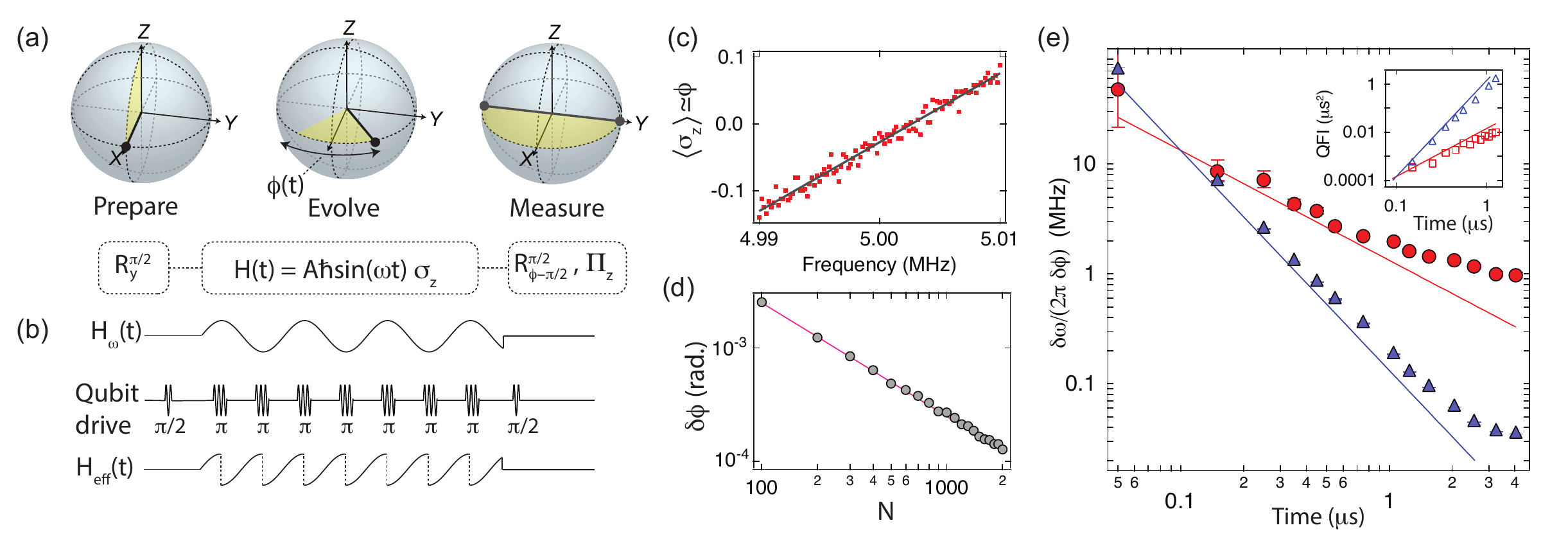}
\end{center}
\caption{\small {\bf Frequency metrology with optimal control.}  (a), Schematic of the estimation task:  The qubit is prepared in a superposition of energy eigenstates $(|0\rangle + |1\rangle)/\sqrt{2}$, followed by an interaction with a time-periodic Hamiltonian with frequency $\omega$ for a certain time, followed a $\pi/2$ pulse and projection in the $\sigma_z$ basis to determine the acquired phase.  (b), The energy eigenvalue difference of Hamiltonian $H_\omega(t)$  is sketched in time, together with the optimal coherent control pulses (repeated $\pi$ pulses at the antinodes of the oscillating Hamiltonian) designed to acquire maximum frequency information. This results in the effective total Hamiltonian, $H_\mathrm{eff}(t)$.  The acquired phase is the time integral of this function.  (c), The frequency sensitivity is determined by varying $\omega$, and a linear fit determines $\mathrm{d}\phi/\mathrm{d}\omega$. (d), The phase uncertainty $\delta \phi$ versus experimental repetition number $N$ shows that the phase uncertainty is given by the binomial error $1/\sqrt{4N}$ (solid line). (e),  The frequency sensitivity, for the uncontrolled (red circles) and optimal control (blue diamonds) attain the respective limits (solid lines) for times shorter than the decoherence time. The error bars indicate the estimated standard deviation of slope $\mathrm{d}\omega/\mathrm{d}\phi$ from the linear regression fit as in panel (c). (e)(Inset), The quantum Fisher information associated with a given measurement protocol (uncontrolled; red, controlled; blue), determined from  the slope of the acquired phase versus frequency is displayed on a log-log plot versus time. }
\label{fig2}
\end{figure*}

To determine the eigenvalues of $h_\omega(T)$, we break the unitary evolution $U_\omega(0\rightarrow T)$ into infinitesimal time intervals as discussed in \cite{pang17}, and consider the eigenvalues of $h_\omega(t)$ versus time. In the current case, the Hamiltonian commutes with itself at different times, so we arrange the system to be in a superposition of the eigenstates of \revkm{ $\partial_\omega H_\omega(t)/\hbar$}, such that the eigenvalues maintain maximal separation.  These eigenvalues  simply evolve as \revkm{$\mu_\pm(t) = \pm  A  t \cos(\omega t)/2$}. In Figure 1b we sketch $\mu_\pm(t)$. The quantum Fisher information about the frequency $\omega$ associated with an evolution for time $T$ is given by,
\begin{align}
I_\omega^{(Q)} = \left[\int_0^T [\mu_+(t) - \mu_-(t)] \mathrm{d}t\right]^2 , \label{eq:iq}
\end{align} which increases as $T^2$.  Figure 1c displays how additional control at the crossing points can be used to dramatically enhance the QFI.  By applying a control to guide the qubit along a trajectory that maximizes the integral (\ref{eq:iq}), the QFI can increase instead as $T^4$ as shown in Figure 1d.  The intuitive reason for the $T^4$ scaling versus the $T^2$ scaling is that for time-independent Hamiltonians, two nearby quantum states corresponding to different values of the parameter can only diverge from each other with constant velocity, whereas in time-dependent Hamiltonians, they can accelerate away from each other, giving greater quantum distinguishability of the states in the same period of time \cite{pang17,yang2016quantum,gefen2017control}.


We now turn to the experiment, where we realize the optimal control depicted in Figure 1c.  The experimental setup consists of a  superconducting transmon circuit \cite{koch07} that is dispersively coupled to a waveguide cavity \cite{paik113D}.  The qubit system is comprised of the lowest two levels of the circuit, and is described by the Pauli spin operators $\sigma_x, \sigma_y, \sigma_z$.  The dispersive interaction between the qubit and the cavity, described by the  Hamiltonian $H_\mathrm{int} = -\hbar \chi \hat{n} \sigma_z$  allows for rapid, quantum non-demolition measurement of the qubit in the energy basis by probing the cavity resonance with microwave photons. Here $\chi/2\pi = -0.5$ MHz is the dispersive coupling rate  and $\hat{n}$ is the cavity photon number operator.  To create the time-dependent Hamiltonian, $H_\omega = A \hbar \sin(\omega t)\sigma_z/2$, we drive the cavity with detuning $\Delta/2\pi = 37$ MHz to populate the cavity with an average $\bar{n} = \bar{n}_0 + A \sin(\omega t)/2\chi$  photons. The mean photon number $\bar{n}_0 = 6.4$ results in an ac Stark shift of $6.4$ MHz and the modulation amplitude $A/2\pi = 0.60$ MHz.





We first demonstrate the standard $T^2$ scaling of the quantum Fisher information that is obtained without Hamiltonian control.  An equal superposition state $(|0\rangle + e^{i\phi} |1\rangle)/\sqrt{2}$ maximizes the QFI, and the measurement protocol is simply a Ramsey sequence as depicted in Figure 2a.  A $\pi/2$ pulse is applied, followed by waiting for a time $T$, followed by a second $\pi/2$ pulse and projective measurement in the $\sigma_z$ basis.   The axis of the second $\pi/2$ rotation is adjusted such that the projective measurement in the energy basis accumulates maximal information about the phase of the qubit.  The QFI is given in terms of the Bures distance \cite{bure69}, $\mathrm{d}s^2 = 2(1-|\langle \psi_{\omega}|\psi_{\omega + \mathrm{d}\omega}\rangle |)$, where $I_\omega^{(Q)} = 4  \mathrm{d}s^2/\mathrm{ d}\omega^2$.  As such, we vary $\omega$ by a small amount to determine the slope (Figure 2c), where $I_\omega^{(Q)} = (\mathrm{d}\phi/\mathrm{d}\omega)^2$. The frequency sensitivity is ultimately governed by the QFI and the phase variance, which as shown in Figure 2d is given by the standard binomial error $\delta \phi = 1/\sqrt{4 N}$ due to projection noise, resulting in a cumulative frequency information of $N I_\omega^{(Q)}$.  As displayed in Figure 2e, the frequency sensitivity improves as $\omega/(A T)$,  (QFI $\propto T^2$) until dephasing of the qubit, characterized by  $T_2^* = 4\ \mu$s degrades the sensitivity.



The key idea behind optimal coherent control is to impose an additional time dependent Hamiltonian $H_\mathrm{c}(t)$ to maximize the difference of the eigenvalues of $h_\omega(T)$. In Figure \ref{fig2}b we display this optimal Hamiltonian control, which consists of discrete unitary $\pi$ rotations applied to the qubit at specific optimal times:  these are applied at the {\it antinodes} of the estimated Hamiltonian, rather that at the nodes as is commonly seen in dynamical decoupling sequences \cite{yang2016quantum}.  In contrast to dynamical decoupling pulses, whose object is to refocus diverging states and prolong coherence, our control pulses do the opposite: the objective is to separate as quickly as possible two quantum states corresponding to nearby values of the frequency in order to improve our resolution of that parameter; hybrid schemes have very recently been proposed \cite{gefen2017control}. In Figure \ref{fig2}e we show how under optimal control the frequency sensitivity attains the ultimate limit $\delta \omega/\delta \phi  = \pi/(A T^2)$ for short times \addkm{and yields better sensitivity over the no-control case as long as $\omega T> \pi$}. This corresponds to a $T^4$ scaling of the QFI. At long times, decoherence of the qubit causes the QFI to decrease due to increasing overlap of the states  $|\psi_{\omega}\rangle$ and $|\psi_{\omega + \mathrm{d}\omega}\rangle$.

The optimal Hamiltonian control yields a $T^2$ improvement over the QFI obtained with a standard Ramsey measurement.  Given a finite time resource in metrology, such as the finite $T_2^*$ time of the qubit, this yields a substantial improvement in QFI, amounting to a factor of  740 in this experimental demonstration.

\addkm{In contrast to recent work \cite{boss17,schm17}  where $T^3$ scaling of the QFI has been observed for times only limited by the stability of an external reference, the $T^4$ scaling observed here is limited to times $T<T_2^*$. If we consider sensing for a duration longer than $T_2$, the optimal approach is to utilize repeated, back-to-back measurements each with duration $T_2$.  By taking advantage of the fact that these repeated measurements sample the signal at different times a $T^3$ scaling of the QFI for the total signal sampling time is also possible with our approach, but with an optimized prefactor.}


Having demonstrated such a significant improvement in the scaling of the quantum Fisher information with time, it is worth inquiring as to whether other Hamiltonian parameters can be estimated with such precision. For example, could the QFI associated with the amplitude $I_A^{(Q)}$ of the time dependent Hamiltonian also achieve such scaling, or at least an improvement under optimal control compared to the uncontrolled case?  To address this, we again consider the eigenvalues of \revkm{$h_A$, $\mu_\pm  = \pm \sin(\omega t)/2$}, which do not increase in time. \addkm{As is well known from work with nitrogen-vacancy spin sensing \cite{dela11,mami13,lore14,sush14,lovc16}, in this case the optimal control strategy is again to apply $\pi$ rotations}, but this time at \emph{nodes} of the Hamiltonian, and yields an overall $T^2$ scaling of the quantum Fisher information as we discuss in the Supplemental Information \revkm{\cite{supp}}. This is an improvement over the no-control case, where the maximum quantum Fisher information does not increase for longer interaction times.

We note that the optimal control needed to obtain the enhanced precision of the frequency depends on knowledge of the phase and frequency, which is itself the parameter to be estimated!  Therefore, in general, we must apply adaptive control \cite{wald12,baum16,yuan15,yuan16} where first some crude knowledge of the parameter is obtained without control, which is then used in the control Hamiltonian to obtain a more precise estimate of the parameter, which is fed back to adjust the coherent control in an adaptive loop until the optimal arrangement is converged upon. One might worry that the $T^4$ scaling is so sensitive to the matching of the time-dependent Hamiltonian and control that the $T^4$ scaling is difficult to achieve in practice.  \addkm{The degradation of the QFI due to frequency mismatch between the control and the parameter was analyzed for the case of a rotating magnetic field in Ref.~\cite{pang17} and by applying a similar analysis here, we find that the QFI in the presence of a frequency mismatch  $\Delta \omega$ is to leading order $I_\omega^{(Q)} = A^2 T^4/\pi^2\ (1-\Delta \omega^2 T^2/2)$. Because the correction grows as $T^2$, an iterative procedure is required to refine the control frequency.  The requirements on matching the phase of the control leads to a correction to the QFI proportional to $(1-\Delta \theta^2)$, which only depends on the phase mismatch, $\Delta \theta$ and does not grow with time \cite{supp}. In Figure~3 we show the experimentally obtained quantum Fisher information for different mismatches between the phase and frequency of the control Hamiltonian.} As shown in Figure 3a, the QFI reaches a maximum when the control is matched to the modulation frequency $\omega$ with vanishing phase offset.  Figure 3a also highlights how  this control landscape can be mapped without knowledge of the parameters that are to be estimated.  

\begin{figure}[h]
\begin{center}
\includegraphics[width = 0.5\textwidth]{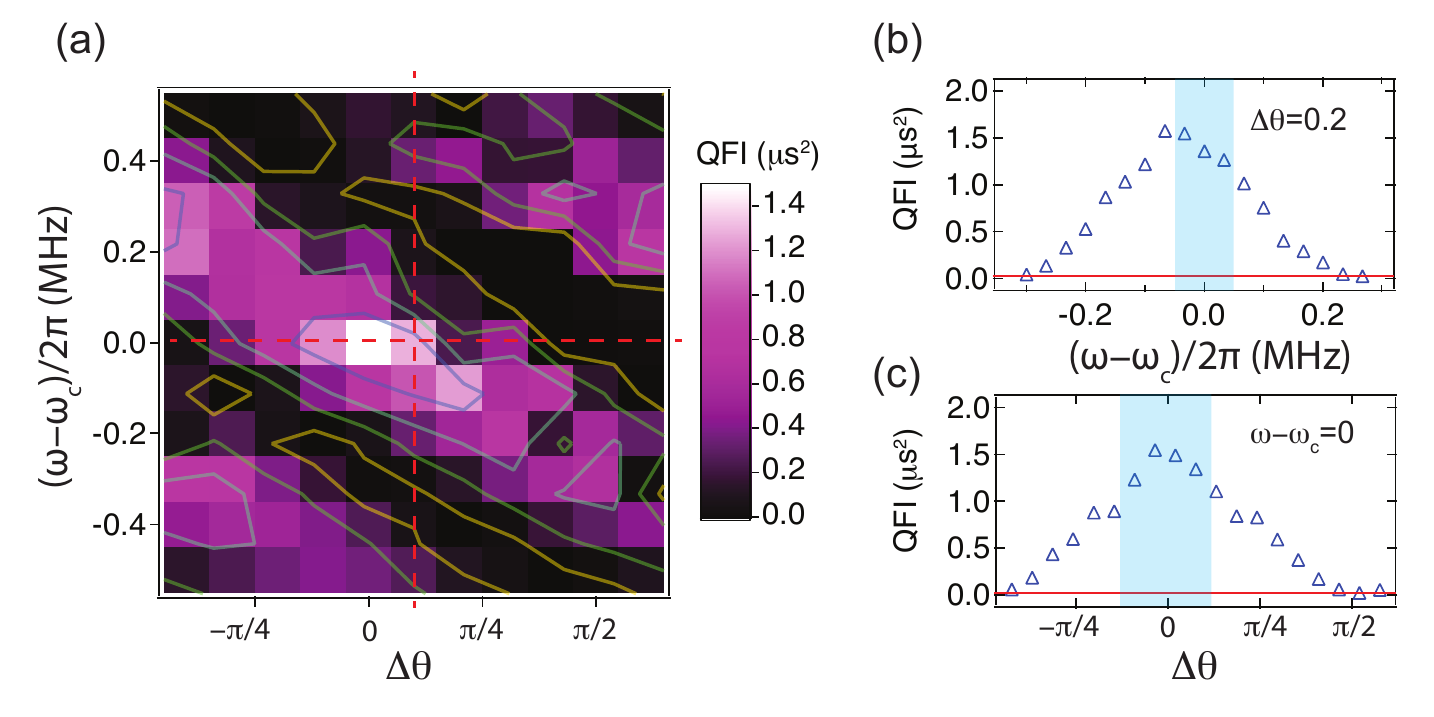}
\end{center}
\caption{\small {\bf Optimal control landscape.}
(a), Color plot of the QFI as two control parameters are swept for $T = 1.25 \ \mu$s:  \revkm{the phase difference between the control $\pi$ pulses the periodic Hamiltonian modulation ($x$-axis), and the duration $T_\mathrm{d}$ between the $\pi$ pulses ($y$-axis) as specified by the control frequency $\omega_\mathrm{c} = 2\pi/T_\mathrm{d}$. The phase difference has been shifted slightly to account for a 6-ns delay between the control pulses and the periodic Hamiltonian. (b,c), Line cuts through the control landscape (locations indicated as dashed red lines in (a))} show that for small parameter mismatches the QFI is still significantly greater than the uncontrolled case (red line). The blue regions show the parameter uncertainty based on uncontrolled estimation using $N=100$ experimental repetitions. }
\label{fig3}
\end{figure}

Figure  3b,c display the QFI versus frequency and phase mismatch in detail. For the $1.25 \ \mu$s interaction time considered here, uncontrolled frequency and phase estimation based on $N=100$ experimental repetitions \revkm{(used to reduce the phase uncertainty)} is sufficient to find the maximum in the QFI available with optimal control. Therefore, the robustness of the optimal control improvement in QFI to variations in the control parameters is sufficient to allow adaptive control to rapidly converge to the optimal values. \revkm{In fact, in Ref.~\cite{pang17} it was proved that the number of iterations required to approach the maximum sensitivity grows only as a double-logarithm of the total time $T$ and in the Supplemental Information \cite{supp} we discuss how an iterative procedure can be used to adaptively improve the frequency precision}.


In quantum enhanced metrology, one seeks to take advantage of quantum properties to maximally utilize the available measurement resources. For parallel resources, such as the number of quantum systems, entanglement can be utilized to achieve Heisenberg scaling.  We have demonstrated how quantum coherence, optimally harnessed through coherent control, can maximally utilize the serial resource: time.  It is theoretically possible to combine both the serial and parallel resources which would give the best case quantum precision.  The advantages conferred in frequency metrology with time-dependent Hamiltonians opens new horizons in precision measurement and time-keeping.

\begin{acknowledgements}
\emph{Acknowledgements}---We acknowledge P. M. Harrington, D. Tan, and J. T. Monroe for discussions, sample fabrication, and contributions to the experimental setup. We acknowledge research support
from NSF grants DMR-1506081 and PHY-1607156, the ONR No.
12114811 and the ARO No. W911NF-15-1-0496. This
research used facilities at the Institute of Materials Science and Engineering at Washington University. K.W.M
acknowledges support from the Sloan Foundation.
\end{acknowledgements}


\begin{thebibliography}{40}
\expandafter\ifx\csname natexlab\endcsname\relax\def\natexlab#1{#1}\fi
\expandafter\ifx\csname bibnamefont\endcsname\relax
  \def\bibnamefont#1{#1}\fi
\expandafter\ifx\csname bibfnamefont\endcsname\relax
  \def\bibfnamefont#1{#1}\fi
\expandafter\ifx\csname citenamefont\endcsname\relax
  \def\citenamefont#1{#1}\fi
\expandafter\ifx\csname url\endcsname\relax
  \def\url#1{\texttt{#1}}\fi
\expandafter\ifx\csname urlprefix\endcsname\relax\def\urlprefix{URL }\fi
\providecommand{\bibinfo}[2]{#2}
\providecommand{\eprint}[2][]{\url{#2}}

\bibitem[{\citenamefont{Abe and et. al.}(2008)}]{Abe2008}
\bibinfo{author}{\bibfnamefont{S.}~\bibnamefont{Abe}} \bibnamefont{and}
  \bibinfo{author}{\bibnamefont{et. al.}} (\bibinfo{collaboration}{The KamLAND
  Collaboration}), \bibinfo{journal}{Phys. Rev. Lett.}
  \textbf{\bibinfo{volume}{100}}, \bibinfo{pages}{221803}
  (\bibinfo{year}{2008}).

\bibitem[{\citenamefont{Walker et~al.}(1994)\citenamefont{Walker, Sheehy,
  DiMauro, Agostini, Schafer, and Kulander}}]{Walk1994}
\bibinfo{author}{\bibfnamefont{B.}~\bibnamefont{Walker}},
  \bibinfo{author}{\bibfnamefont{B.}~\bibnamefont{Sheehy}},
  \bibinfo{author}{\bibfnamefont{L.~F.} \bibnamefont{DiMauro}},
  \bibinfo{author}{\bibfnamefont{P.}~\bibnamefont{Agostini}},
  \bibinfo{author}{\bibfnamefont{K.~J.} \bibnamefont{Schafer}},
  \bibnamefont{and} \bibinfo{author}{\bibfnamefont{K.~C.}
  \bibnamefont{Kulander}}, \bibinfo{journal}{Phys. Rev. Lett.}
  \textbf{\bibinfo{volume}{73}}, \bibinfo{pages}{1227} (\bibinfo{year}{1994}).

\bibitem[{\citenamefont{Aguilar and et. al.}(2013)}]{Agui2013}
\bibinfo{author}{\bibfnamefont{M.}~\bibnamefont{Aguilar}} \bibnamefont{and}
  \bibinfo{author}{\bibnamefont{et. al.}} (\bibinfo{collaboration}{AMS
  Collaboration}), \bibinfo{journal}{Phys. Rev. Lett.}
  \textbf{\bibinfo{volume}{110}}, \bibinfo{pages}{141102}
  (\bibinfo{year}{2013}).

\bibitem[{\citenamefont{Rosi et~al.}(2014)\citenamefont{Rosi, Sorrentino,
  Cacciapuoti, Prevedelli, and Tino}}]{rosi2014}
\bibinfo{author}{\bibfnamefont{G.}~\bibnamefont{Rosi}},
  \bibinfo{author}{\bibfnamefont{F.}~\bibnamefont{Sorrentino}},
  \bibinfo{author}{\bibfnamefont{L.}~\bibnamefont{Cacciapuoti}},
  \bibinfo{author}{\bibfnamefont{M.}~\bibnamefont{Prevedelli}},
  \bibnamefont{and} \bibinfo{author}{\bibfnamefont{G.}~\bibnamefont{Tino}},
  \bibinfo{journal}{Nature} \textbf{\bibinfo{volume}{510}},
  \bibinfo{pages}{518} (\bibinfo{year}{2014}).

\bibitem[{\citenamefont{Essen and Parry}(1955)}]{essen1955atomic}
\bibinfo{author}{\bibfnamefont{L.}~\bibnamefont{Essen}} \bibnamefont{and}
  \bibinfo{author}{\bibfnamefont{J.}~\bibnamefont{Parry}},
  \bibinfo{journal}{Nature} \textbf{\bibinfo{volume}{176}},
  \bibinfo{pages}{280} (\bibinfo{year}{1955}).

\bibitem[{\citenamefont{Lyons}(1957)}]{lyons1957atomic}
\bibinfo{author}{\bibfnamefont{H.}~\bibnamefont{Lyons}},
  \bibinfo{journal}{Scientific American} \textbf{\bibinfo{volume}{196}},
  \bibinfo{pages}{71} (\bibinfo{year}{1957}).

\bibitem[{\citenamefont{Ludlow et~al.}(2015)\citenamefont{Ludlow, Boyd, Ye,
  Peik, and Schmidt}}]{ludl15rmp}
\bibinfo{author}{\bibfnamefont{A.~D.} \bibnamefont{Ludlow}},
  \bibinfo{author}{\bibfnamefont{M.~M.} \bibnamefont{Boyd}},
  \bibinfo{author}{\bibfnamefont{J.}~\bibnamefont{Ye}},
  \bibinfo{author}{\bibfnamefont{E.}~\bibnamefont{Peik}}, \bibnamefont{and}
  \bibinfo{author}{\bibfnamefont{P.~O.} \bibnamefont{Schmidt}},
  \bibinfo{journal}{Rev. Mod. Phys.} \textbf{\bibinfo{volume}{87}},
  \bibinfo{pages}{637} (\bibinfo{year}{2015}).

\bibitem[{\citenamefont{Gabor}(1946)}]{gabor1946theory}
\bibinfo{author}{\bibfnamefont{D.}~\bibnamefont{Gabor}},
  \bibinfo{journal}{Journal of the Institution of Electrical Engineers-Part
  III: Radio and Communication Engineering} \textbf{\bibinfo{volume}{93}},
  \bibinfo{pages}{429} (\bibinfo{year}{1946}).

\bibitem[{\citenamefont{Cram\'er}(1946)}]{cram46}
\bibinfo{author}{\bibfnamefont{H.}~\bibnamefont{Cram\'er}},
  \emph{\bibinfo{title}{Mathematical Methods of Statistics}}
  (\bibinfo{publisher}{Princeton University Press}, \bibinfo{year}{1946}).

\bibitem[{\citenamefont{Fisher}(1925)}]{fish25}
\bibinfo{author}{\bibfnamefont{R.~A.} \bibnamefont{Fisher}},
  \bibinfo{journal}{Math. Proc. Cambridge Philos. Soc.}
  \textbf{\bibinfo{volume}{22}}, \bibinfo{pages}{700} (\bibinfo{year}{1925}).

\bibitem[{\citenamefont{Braunstein and Caves}(1994)}]{Brau94}
\bibinfo{author}{\bibfnamefont{S.~L.} \bibnamefont{Braunstein}}
  \bibnamefont{and} \bibinfo{author}{\bibfnamefont{C.~M.} \bibnamefont{Caves}},
  \bibinfo{journal}{Phys. Rev. Lett.} \textbf{\bibinfo{volume}{72}},
  \bibinfo{pages}{3439} (\bibinfo{year}{1994}).

\bibitem[{\citenamefont{Braunstein et~al.}(1996)\citenamefont{Braunstein,
  Caves, and Milburn}}]{Brau96}
\bibinfo{author}{\bibfnamefont{S.~L.} \bibnamefont{Braunstein}},
  \bibinfo{author}{\bibfnamefont{C.~M.} \bibnamefont{Caves}}, \bibnamefont{and}
  \bibinfo{author}{\bibfnamefont{G.}~\bibnamefont{Milburn}},
  \bibinfo{journal}{Annals of Physics} \textbf{\bibinfo{volume}{247}},
  \bibinfo{pages}{135 } (\bibinfo{year}{1996}), ISSN \bibinfo{issn}{0003-4916}.

\bibitem[{\citenamefont{Alipour et~al.}(2014)\citenamefont{Alipour, Mehboudi,
  and Rezakhani}}]{Alip2014}
\bibinfo{author}{\bibfnamefont{S.}~\bibnamefont{Alipour}},
  \bibinfo{author}{\bibfnamefont{M.}~\bibnamefont{Mehboudi}}, \bibnamefont{and}
  \bibinfo{author}{\bibfnamefont{A.~T.} \bibnamefont{Rezakhani}},
  \bibinfo{journal}{Phys. Rev. Lett.} \textbf{\bibinfo{volume}{112}},
  \bibinfo{pages}{120405} (\bibinfo{year}{2014}).

\bibitem[{\citenamefont{Tsang et~al.}(2011)\citenamefont{Tsang, Wiseman, and
  Caves}}]{Tsan2011}
\bibinfo{author}{\bibfnamefont{M.}~\bibnamefont{Tsang}},
  \bibinfo{author}{\bibfnamefont{H.~M.} \bibnamefont{Wiseman}},
  \bibnamefont{and} \bibinfo{author}{\bibfnamefont{C.~M.} \bibnamefont{Caves}},
  \bibinfo{journal}{Phys. Rev. Lett.} \textbf{\bibinfo{volume}{106}},
  \bibinfo{pages}{090401} (\bibinfo{year}{2011}).

\bibitem[{\citenamefont{Giovannetti et~al.}(2004)\citenamefont{Giovannetti,
  Lloyd, and Maccone}}]{giov04}
\bibinfo{author}{\bibfnamefont{V.}~\bibnamefont{Giovannetti}},
  \bibinfo{author}{\bibfnamefont{S.}~\bibnamefont{Lloyd}}, \bibnamefont{and}
  \bibinfo{author}{\bibfnamefont{L.}~\bibnamefont{Maccone}},
  \textbf{\bibinfo{volume}{306}}, \bibinfo{pages}{1330} (\bibinfo{year}{2004}),
  ISSN \bibinfo{issn}{0036-8075}.

\bibitem[{\citenamefont{Kolobov}(1999)}]{kolo99}
\bibinfo{author}{\bibfnamefont{M.~I.} \bibnamefont{Kolobov}},
  \bibinfo{journal}{Rev. Mod. Phys.} \textbf{\bibinfo{volume}{71}},
  \bibinfo{pages}{1539} (\bibinfo{year}{1999}).

\bibitem[{\citenamefont{Collaboration}(2011)}]{ligo11}
\bibinfo{author}{\bibfnamefont{T.~L.~S.} \bibnamefont{Collaboration}},
  \bibinfo{journal}{Nature Physics} \textbf{\bibinfo{volume}{7}},
  \bibinfo{pages}{962} (\bibinfo{year}{2011}).

\bibitem[{\citenamefont{Sewell et~al.}(2012)\citenamefont{Sewell, Koschorreck,
  Napolitano, Dubost, Behbood, and Mitchell}}]{sewe12}
\bibinfo{author}{\bibfnamefont{R.~J.} \bibnamefont{Sewell}},
  \bibinfo{author}{\bibfnamefont{M.}~\bibnamefont{Koschorreck}},
  \bibinfo{author}{\bibfnamefont{M.}~\bibnamefont{Napolitano}},
  \bibinfo{author}{\bibfnamefont{B.}~\bibnamefont{Dubost}},
  \bibinfo{author}{\bibfnamefont{N.}~\bibnamefont{Behbood}}, \bibnamefont{and}
  \bibinfo{author}{\bibfnamefont{M.~W.} \bibnamefont{Mitchell}},
  \bibinfo{journal}{Phys. Rev. Lett.} \textbf{\bibinfo{volume}{109}},
  \bibinfo{pages}{253605} (\bibinfo{year}{2012}).

\bibitem[{\citenamefont{Itano et~al.}(1993)\citenamefont{Itano, Bergquist,
  Bollinger, Gilligan, Heinzen, Moore, Raizen, and Wineland}}]{itan93}
\bibinfo{author}{\bibfnamefont{W.~M.} \bibnamefont{Itano}},
  \bibinfo{author}{\bibfnamefont{J.~C.} \bibnamefont{Bergquist}},
  \bibinfo{author}{\bibfnamefont{J.~J.} \bibnamefont{Bollinger}},
  \bibinfo{author}{\bibfnamefont{J.~M.} \bibnamefont{Gilligan}},
  \bibinfo{author}{\bibfnamefont{D.~J.} \bibnamefont{Heinzen}},
  \bibinfo{author}{\bibfnamefont{F.~L.} \bibnamefont{Moore}},
  \bibinfo{author}{\bibfnamefont{M.~G.} \bibnamefont{Raizen}},
  \bibnamefont{and} \bibinfo{author}{\bibfnamefont{D.~J.}
  \bibnamefont{Wineland}}, \bibinfo{journal}{Phys. Rev. A}
  \textbf{\bibinfo{volume}{47}}, \bibinfo{pages}{3554} (\bibinfo{year}{1993}).

\bibitem[{\citenamefont{Giovannetti et~al.}(2006)\citenamefont{Giovannetti,
  Lloyd, and Maccone}}]{giov06}
\bibinfo{author}{\bibfnamefont{V.}~\bibnamefont{Giovannetti}},
  \bibinfo{author}{\bibfnamefont{S.}~\bibnamefont{Lloyd}}, \bibnamefont{and}
  \bibinfo{author}{\bibfnamefont{L.}~\bibnamefont{Maccone}},
  \bibinfo{journal}{Phys. Rev. Lett.} \textbf{\bibinfo{volume}{96}},
  \bibinfo{pages}{010401} (\bibinfo{year}{2006}).

\bibitem[{\citenamefont{de~Clercq et~al.}(2016)\citenamefont{de~Clercq, Oswald,
  Fl\"uhmann, Keitch, Kienzler, Lo, Marinelli, Nadlinger, Negnevitsky, and
  Home}}]{cler15}
\bibinfo{author}{\bibfnamefont{L.}~\bibnamefont{de~Clercq}},
  \bibinfo{author}{\bibfnamefont{R.}~\bibnamefont{Oswald}},
  \bibinfo{author}{\bibfnamefont{C.}~\bibnamefont{Fl\"uhmann}},
  \bibinfo{author}{\bibfnamefont{B.}~\bibnamefont{Keitch}},
  \bibinfo{author}{\bibfnamefont{D.}~\bibnamefont{Kienzler}},
  \bibinfo{author}{\bibfnamefont{H.-Y.} \bibnamefont{Lo}},
  \bibinfo{author}{\bibfnamefont{M.}~\bibnamefont{Marinelli}},
  \bibinfo{author}{\bibfnamefont{D.}~\bibnamefont{Nadlinger}},
  \bibinfo{author}{\bibfnamefont{V.}~\bibnamefont{Negnevitsky}},
  \bibnamefont{and} \bibinfo{author}{\bibfnamefont{J.}~\bibnamefont{Home}},
  \bibinfo{journal}{Nat. Commun.} \textbf{\bibinfo{volume}{7}},
  \bibinfo{pages}{11218} (\bibinfo{year}{2016}).

\bibitem[{\citenamefont{Pang and Jordan}(2017)}]{pang17}
\bibinfo{author}{\bibfnamefont{S.}~\bibnamefont{Pang}} \bibnamefont{and}
  \bibinfo{author}{\bibfnamefont{A.~N.} \bibnamefont{Jordan}},
  \bibinfo{journal}{Nat. Commun.} \textbf{\bibinfo{volume}{8}},
  \bibinfo{pages}{14695} (\bibinfo{year}{2017}).

\bibitem[{\citenamefont{Boss et~al.}(2017)\citenamefont{Boss, Cujia, Zopes, and
  Degen}}]{boss17}
\bibinfo{author}{\bibfnamefont{J.~M.} \bibnamefont{Boss}},
  \bibinfo{author}{\bibfnamefont{K.~S.} \bibnamefont{Cujia}},
  \bibinfo{author}{\bibfnamefont{J.}~\bibnamefont{Zopes}}, \bibnamefont{and}
  \bibinfo{author}{\bibfnamefont{C.~L.} \bibnamefont{Degen}},
  \bibinfo{journal}{Science} \textbf{\bibinfo{volume}{356}},
  \bibinfo{pages}{837} (\bibinfo{year}{2017}).

\bibitem[{\citenamefont{Schmitt et~al.}(2017)\citenamefont{Schmitt, Gefen,
  St{\"u}rner, Unden, Wolff, M{\"u}ller, Scheuer, Naydenov, Markham, Pezzagna
  et~al.}}]{schm17}
\bibinfo{author}{\bibfnamefont{S.}~\bibnamefont{Schmitt}},
  \bibinfo{author}{\bibfnamefont{T.}~\bibnamefont{Gefen}},
  \bibinfo{author}{\bibfnamefont{F.~M.} \bibnamefont{St{\"u}rner}},
  \bibinfo{author}{\bibfnamefont{T.}~\bibnamefont{Unden}},
  \bibinfo{author}{\bibfnamefont{G.}~\bibnamefont{Wolff}},
  \bibinfo{author}{\bibfnamefont{C.}~\bibnamefont{M{\"u}ller}},
  \bibinfo{author}{\bibfnamefont{J.}~\bibnamefont{Scheuer}},
  \bibinfo{author}{\bibfnamefont{B.}~\bibnamefont{Naydenov}},
  \bibinfo{author}{\bibfnamefont{M.}~\bibnamefont{Markham}},
  \bibinfo{author}{\bibfnamefont{S.}~\bibnamefont{Pezzagna}},
  \bibnamefont{et~al.}, \bibinfo{journal}{Science}
  \textbf{\bibinfo{volume}{356}}, \bibinfo{pages}{832} (\bibinfo{year}{2017}).

\bibitem[{\citenamefont{Jordan}(2017)}]{jord17}
\bibinfo{author}{\bibfnamefont{A.~N.} \bibnamefont{Jordan}},
  \bibinfo{journal}{Science} \textbf{\bibinfo{volume}{356}},
  \bibinfo{pages}{802} (\bibinfo{year}{2017}).

\bibitem[{\citenamefont{Yang et~al.}(2017)\citenamefont{Yang, Pang, and
  Jordan}}]{yang2016quantum}
\bibinfo{author}{\bibfnamefont{J.}~\bibnamefont{Yang}},
  \bibinfo{author}{\bibfnamefont{S.}~\bibnamefont{Pang}}, \bibnamefont{and}
  \bibinfo{author}{\bibfnamefont{A.~N.} \bibnamefont{Jordan}},
  \bibinfo{journal}{Phys. Rev. A} \textbf{\bibinfo{volume}{96}},
  \bibinfo{pages}{020301} (\bibinfo{year}{2017}).

\bibitem[{\citenamefont{Gefen et~al.}(2017)\citenamefont{Gefen, Jelezko, and
  Retzker}}]{gefen2017control}
\bibinfo{author}{\bibfnamefont{T.}~\bibnamefont{Gefen}},
  \bibinfo{author}{\bibfnamefont{F.}~\bibnamefont{Jelezko}}, \bibnamefont{and}
  \bibinfo{author}{\bibfnamefont{A.}~\bibnamefont{Retzker}},
  \bibinfo{journal}{arXiv preprint arXiv:1702.07408}  (\bibinfo{year}{2017}).

\bibitem[{\citenamefont{Koch et~al.}(2007)\citenamefont{Koch, Yu, Gambetta,
  Houck, Schuster, Majer, Blais, Devoret, Girvin, and Schoelkopf}}]{koch07}
\bibinfo{author}{\bibfnamefont{J.}~\bibnamefont{Koch}},
  \bibinfo{author}{\bibfnamefont{T.~M.} \bibnamefont{Yu}},
  \bibinfo{author}{\bibfnamefont{J.}~\bibnamefont{Gambetta}},
  \bibinfo{author}{\bibfnamefont{A.~A.} \bibnamefont{Houck}},
  \bibinfo{author}{\bibfnamefont{D.~I.} \bibnamefont{Schuster}},
  \bibinfo{author}{\bibfnamefont{J.}~\bibnamefont{Majer}},
  \bibinfo{author}{\bibfnamefont{A.}~\bibnamefont{Blais}},
  \bibinfo{author}{\bibfnamefont{M.~H.} \bibnamefont{Devoret}},
  \bibinfo{author}{\bibfnamefont{S.~M.} \bibnamefont{Girvin}},
  \bibnamefont{and} \bibinfo{author}{\bibfnamefont{R.~J.}
  \bibnamefont{Schoelkopf}}, \bibinfo{journal}{Phys. Rev. A}
  \textbf{\bibinfo{volume}{76}}, \bibinfo{pages}{042319}
  (\bibinfo{year}{2007}).

\bibitem[{\citenamefont{Paik et~al.}(2011)\citenamefont{Paik, Schuster, Bishop,
  Kirchmair, Catelani, Sears, Johnson, Reagor, Frunzio, Glazman
  et~al.}}]{paik113D}
\bibinfo{author}{\bibfnamefont{H.}~\bibnamefont{Paik}},
  \bibinfo{author}{\bibfnamefont{D.~I.} \bibnamefont{Schuster}},
  \bibinfo{author}{\bibfnamefont{L.~S.} \bibnamefont{Bishop}},
  \bibinfo{author}{\bibfnamefont{G.}~\bibnamefont{Kirchmair}},
  \bibinfo{author}{\bibfnamefont{G.}~\bibnamefont{Catelani}},
  \bibinfo{author}{\bibfnamefont{A.~P.} \bibnamefont{Sears}},
  \bibinfo{author}{\bibfnamefont{B.~R.} \bibnamefont{Johnson}},
  \bibinfo{author}{\bibfnamefont{M.~J.} \bibnamefont{Reagor}},
  \bibinfo{author}{\bibfnamefont{L.}~\bibnamefont{Frunzio}},
  \bibinfo{author}{\bibfnamefont{L.~I.} \bibnamefont{Glazman}},
  \bibnamefont{et~al.}, \bibinfo{journal}{Phys. Rev. Lett.}
  \textbf{\bibinfo{volume}{107}}, \bibinfo{pages}{240501}
  (\bibinfo{year}{2011}).

\bibitem[{\citenamefont{Bures}(1969)}]{bure69}
\bibinfo{author}{\bibfnamefont{D.}~\bibnamefont{Bures}},
  \bibinfo{journal}{Transactions of the American Mathematical Society}
  \textbf{\bibinfo{volume}{135}}, \bibinfo{pages}{199} (\bibinfo{year}{1969}).

\bibitem[{\citenamefont{de~Lange et~al.}(2011)\citenamefont{de~Lange, Rist\`e,
  Dobrovitski, and Hanson}}]{dela11}
\bibinfo{author}{\bibfnamefont{G.}~\bibnamefont{de~Lange}},
  \bibinfo{author}{\bibfnamefont{D.}~\bibnamefont{Rist\`e}},
  \bibinfo{author}{\bibfnamefont{V.~V.} \bibnamefont{Dobrovitski}},
  \bibnamefont{and} \bibinfo{author}{\bibfnamefont{R.}~\bibnamefont{Hanson}},
  \bibinfo{journal}{Phys. Rev. Lett.} \textbf{\bibinfo{volume}{106}},
  \bibinfo{pages}{080802} (\bibinfo{year}{2011}).

\bibitem[{\citenamefont{Mamin et~al.}(2013)\citenamefont{Mamin, Kim, Sherwood,
  Rettner, Ohno, Awschalom, and Rugar}}]{mami13}
\bibinfo{author}{\bibfnamefont{H.~J.} \bibnamefont{Mamin}},
  \bibinfo{author}{\bibfnamefont{M.}~\bibnamefont{Kim}},
  \bibinfo{author}{\bibfnamefont{M.~H.} \bibnamefont{Sherwood}},
  \bibinfo{author}{\bibfnamefont{C.~T.} \bibnamefont{Rettner}},
  \bibinfo{author}{\bibfnamefont{K.}~\bibnamefont{Ohno}},
  \bibinfo{author}{\bibfnamefont{D.~D.} \bibnamefont{Awschalom}},
  \bibnamefont{and} \bibinfo{author}{\bibfnamefont{D.}~\bibnamefont{Rugar}},
  \bibinfo{journal}{Science} \textbf{\bibinfo{volume}{339}},
  \bibinfo{pages}{557} (\bibinfo{year}{2013}).

\bibitem[{\citenamefont{Loretz et~al.}(2014)\citenamefont{Loretz, Pezzagna,
  Meijer, and Degen}}]{lore14}
\bibinfo{author}{\bibfnamefont{M.}~\bibnamefont{Loretz}},
  \bibinfo{author}{\bibfnamefont{S.}~\bibnamefont{Pezzagna}},
  \bibinfo{author}{\bibfnamefont{J.}~\bibnamefont{Meijer}}, \bibnamefont{and}
  \bibinfo{author}{\bibfnamefont{C.~L.} \bibnamefont{Degen}},
  \bibinfo{journal}{Applied Physics Letters} \textbf{\bibinfo{volume}{104}},
  \bibinfo{pages}{033102} (\bibinfo{year}{2014}).

\bibitem[{\citenamefont{Sushkov et~al.}(2014)\citenamefont{Sushkov, Lovchinsky,
  Chisholm, Walsworth, Park, and Lukin}}]{sush14}
\bibinfo{author}{\bibfnamefont{A.~O.} \bibnamefont{Sushkov}},
  \bibinfo{author}{\bibfnamefont{I.}~\bibnamefont{Lovchinsky}},
  \bibinfo{author}{\bibfnamefont{N.}~\bibnamefont{Chisholm}},
  \bibinfo{author}{\bibfnamefont{R.~L.} \bibnamefont{Walsworth}},
  \bibinfo{author}{\bibfnamefont{H.}~\bibnamefont{Park}}, \bibnamefont{and}
  \bibinfo{author}{\bibfnamefont{M.~D.} \bibnamefont{Lukin}},
  \bibinfo{journal}{Phys. Rev. Lett.} \textbf{\bibinfo{volume}{113}},
  \bibinfo{pages}{197601} (\bibinfo{year}{2014}).

\bibitem[{\citenamefont{Lovchinsky et~al.}(2016)\citenamefont{Lovchinsky,
  Sushkov, Urbach, de~Leon, Choi, De~Greve, Evans, Gertner, Bersin, M{\"u}ller
  et~al.}}]{lovc16}
\bibinfo{author}{\bibfnamefont{I.}~\bibnamefont{Lovchinsky}},
  \bibinfo{author}{\bibfnamefont{A.~O.} \bibnamefont{Sushkov}},
  \bibinfo{author}{\bibfnamefont{E.}~\bibnamefont{Urbach}},
  \bibinfo{author}{\bibfnamefont{N.~P.} \bibnamefont{de~Leon}},
  \bibinfo{author}{\bibfnamefont{S.}~\bibnamefont{Choi}},
  \bibinfo{author}{\bibfnamefont{K.}~\bibnamefont{De~Greve}},
  \bibinfo{author}{\bibfnamefont{R.}~\bibnamefont{Evans}},
  \bibinfo{author}{\bibfnamefont{R.}~\bibnamefont{Gertner}},
  \bibinfo{author}{\bibfnamefont{E.}~\bibnamefont{Bersin}},
  \bibinfo{author}{\bibfnamefont{C.}~\bibnamefont{M{\"u}ller}},
  \bibnamefont{et~al.}, \bibinfo{journal}{Science}
  \textbf{\bibinfo{volume}{351}}, \bibinfo{pages}{836} (\bibinfo{year}{2016}).

\bibitem[{sup()}]{supp}
\bibinfo{note}{Supplementary file}.

\bibitem[{\citenamefont{Waldherr et~al.}(2012)\citenamefont{Waldherr, Beck,
  Neumann, Said, Nitsche, Markham, Twitchen, Twamley, Jelezko, and
  Wrachtrup}}]{wald12}
\bibinfo{author}{\bibfnamefont{G.}~\bibnamefont{Waldherr}},
  \bibinfo{author}{\bibfnamefont{J.}~\bibnamefont{Beck}},
  \bibinfo{author}{\bibfnamefont{P.}~\bibnamefont{Neumann}},
  \bibinfo{author}{\bibfnamefont{R.~S.} \bibnamefont{Said}},
  \bibinfo{author}{\bibfnamefont{M.}~\bibnamefont{Nitsche}},
  \bibinfo{author}{\bibfnamefont{M.~L.} \bibnamefont{Markham}},
  \bibinfo{author}{\bibfnamefont{D.~J.} \bibnamefont{Twitchen}},
  \bibinfo{author}{\bibfnamefont{J.}~\bibnamefont{Twamley}},
  \bibinfo{author}{\bibfnamefont{F.}~\bibnamefont{Jelezko}}, \bibnamefont{and}
  \bibinfo{author}{\bibfnamefont{J.}~\bibnamefont{Wrachtrup}},
  \bibinfo{journal}{Nature Nanotechnology} \textbf{\bibinfo{volume}{7}},
  \bibinfo{pages}{105} (\bibinfo{year}{2012}).

\bibitem[{\citenamefont{Baumgratz and Datta}(2016)}]{baum16}
\bibinfo{author}{\bibfnamefont{T.}~\bibnamefont{Baumgratz}} \bibnamefont{and}
  \bibinfo{author}{\bibfnamefont{A.}~\bibnamefont{Datta}},
  \bibinfo{journal}{Phys. Rev. Lett.} \textbf{\bibinfo{volume}{116}},
  \bibinfo{pages}{030801} (\bibinfo{year}{2016}).

\bibitem[{\citenamefont{Yuan and Fung}(2015)}]{yuan15}
\bibinfo{author}{\bibfnamefont{H.}~\bibnamefont{Yuan}} \bibnamefont{and}
  \bibinfo{author}{\bibfnamefont{C.-H.~F.} \bibnamefont{Fung}},
  \bibinfo{journal}{Phys. Rev. Lett.} \textbf{\bibinfo{volume}{115}},
  \bibinfo{pages}{110401} (\bibinfo{year}{2015}).

\bibitem[{\citenamefont{Yuan}(2016)}]{yuan16}
\bibinfo{author}{\bibfnamefont{H.}~\bibnamefont{Yuan}}, \bibinfo{journal}{Phys.
  Rev. Lett.} \textbf{\bibinfo{volume}{117}}, \bibinfo{pages}{160801}
  (\bibinfo{year}{2016}).

\end{thebibliography}

\appendix

\pagebreak

\textcolor{white}{blank page}
\pagebreak

\onecolumngrid

\section*{Supplemental Information}

\setcounter{figure}{0}

This document contains \revkm{six} sections that provide  supplemental information  to the  data presented in the main text. Section I provides further details on the qubit sample  and  experimental setup.  Section II  presents the  QFI associated with amplitude estimation  discussed in the main text. Section III  analyzes how the QFI scaling depends on the phase mismatch between the signal and the control.  \revkm{Section IV discusses an iterative procedure for improving the frequency precision and the scaling of the QFI on the measurement time. Section V provides and analysis and comparison to the technique of Rabi spectroscopy. Section VI provides statistical analysis used in the main text and supplemental figures as well as specific parameters used in each graph.}
\\

\noindent{\bf I. Experimental setup and sample parameters.}

The experiment  utilizes a single transmon circuit  with a qubit transition  frequency of  $f_\mathrm{q}  = 5.07$ GHz  which is dispersively coupled to a waveguide cavity with resonance  frequency $f_\mathrm{c} = 6.667$ GHz.  The  coherence properties $T_1 = 14 \ \mu$s and $T_2^* =4 \ \mu$s are measured with standard techniques.  The time-dependent Hamiltonian is imposed via an ac Stark shift, which does not degrade the qubit coherence properties, owing to the substantial detuning of the drive.  

Qubit state readout is performed by  resonantly driving the cavity for $500$ ns and amplifying the  transmitted  microwave signal with a  phase sensitive Josephson amplifier.  The combined readout and state preparation fidelity of $80\%$ is determined by the contrast of Rabi oscillations.  The measured  sensitivities are not  corrected for this fidelity.

The  control Hamiltonian consists of $10$ ns duration $\pi$ rotations. All qubit drives are applied  using single sideband modulation  with an offset frequency of $150$ MHz with the carrier situated below the qubit transition frequency. \revkm{The total effective Hamiltonian including the control is given by, 
\begin{align}
H = -\hbar \omega_\mathrm{q} \sigma_z/2 + A \hbar \sin(\omega t)\sigma_z/2 + \hbar f_\mathrm{c}(t) \sigma_y/2,
\end{align}
Where the control $f_\mathrm{c}(t)$ describes repeated 10-ns square pulses of amplitude $\Omega_R/2\pi = 50$ MHz centered at times $t = (2 n +1)\pi/(2\omega)$ with $n = 0, 1, 2,\ldots$ giving the number of control pulses.}
\\

\noindent{\bf II.  Optimal control over amplitude estimation.}

\begin{figure*}
\renewcommand\figurename{Supplemental Figure}
\begin{center}
\includegraphics[width = .4\textwidth]{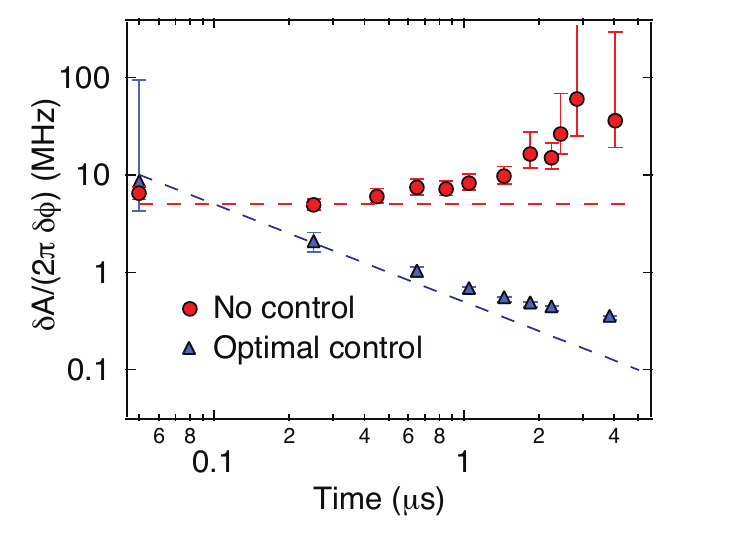}
\end{center}
\caption{\small {\bf Amplitude estimation with optimal control.} The amplitude sensitivity versus interaction time for the uncontrolled (red circles) and optimal control (blue triangles). Whereas the sensitivity does not improve with longer interaction time for the uncontrolled case, optimal control allows the frequency sensitivity to approach the fundamental bound $\delta A  = \pi/T $.}
\label{fig:amp}
\end{figure*}

As discussed in the main text, optimal control can be used to improve the  QFI associated with amplitude estimation; whereas the in the  uncontrolled case, the QFI associated with the amplitude does not increase in time, optimal control can saturate the $T^2$ bound in this case. The respective sensitivities can be determined from the eigenvalues of \revkm{$\partial_A H_A(t)/\hbar$}, which are simply \addkm{$\mu_\pm = \pm \sin(\omega t)/2$}.  Therefore the maximum QFI without control (Eq.\ \ref{eq:iq}) is at most $I_A^{(Q)} =1/\omega^2$, corresponding to an amplitude sensitivity $\delta A/\delta \phi =\omega$.  Under optimal control the QFI can improve in time as $T^2/\pi^2$.      Supplemental Figure \ref{fig:amp} displays the measured frequency sensitivity which is determined from the slope $\mathrm{d}\phi/\mathrm{d}A$ which is determined from a $7\%$ variation in $A$.  The Figure shows that both amplitude measurements are in agreement with their respective bounds.\\

\noindent {\bf III. \addkm{Dependence of the quantum Fisher information on the phase mismatch.}}

\addkm{To study the scaling of the QFI with time in the presence of a phase mismatch between the control and the signal we consider a measurement sequence that contains $N$ oscillatory cycles of the Hamiltonian and a phase mismatch $\Delta \theta$.  In this case the $\pi$-pulses are not applied at the optimal times, shifting the periodic integral given by Eq. \ref{eq:iq}, leading to a QFI that scales as,
\begin{align}
I_\omega^Q = \left[\pi N^2 A \frac{\cos \Delta\theta}{\omega^2} + 2 N \frac{\Delta\phi  \cos \Delta\theta - \sin \Delta \theta}{\omega^2}\right]^2
\end{align}  
For large $N$, the modification to the QFI is given by the first term, which for small $\Delta \theta$ leads to a $(1-\Delta\theta^2)$ reduction in the QFI. }\\

\noindent {\bf IV. Iterative scaling of precision.}

 When we carry out the iterative improvement of Fisher information via adaptive updating of the estimated frequency in the control Hamiltonian, the total time is not typically equally distributed.  Instead, the later iterations will take much longer than the earlier ones, and increase rapidly, so the last few iterations will take nearly all of the total time, as a percentage.
 
 Suppose we make a first crude measurement, and obtain a QFI of $I_0$, with a repetition of this measurement for $N$ times, so the total frequency uncertainty is bounded by
 \be
 \la \delta \omega^2 \ra \ge 1/(N I_0).
 \ee
 When we add control with the frequency mismatch set equal to the standard deviation of the first measurement, and chose the set the free evolution time to be $T_1 = \sqrt{I_0}$, we find for our protocol the QFI for this iteration to be
 \be
 I_1 = \frac{A^2 T_1^4}{\pi^2} (1 - \delta \omega^2 T_1^2/2) =  \frac{A^2 T_1^4}{\pi^2} (1 - \frac{T_1^2}{2 N I_0}).
 \ee
 This measurement (set at time $T_1$) is then itself repeated $N$ times to get a more refined estimate of the frequency.  The next step is to set a longer time, $T_2 = \sqrt{I_1}$, and so on.  Setting the $n^{th}$ time step to be $T_n = \sqrt{I_{n-1}}$, we find
 \be
 I_n = I_0^{2^n} \left( \frac{A}{\pi}^2 (1-1/(2N))\right)^{2^n-1},
 \ee
 so the information scales as a double exponential with the number of iterations.
 
 From this, we can easily see that the number of iterations needed to achieve a given duration $T$ is given when $T_n = T$, or $n$ scales as a double logarithm of the duration $T$.  The total time is given by $N$ times the sum of all $T_n$ from $n=0$ up to the desired number of iterations.  This sum over $n$, in the exponent of the exponent has no analytic solution, so we can take two limits to bound the time.  The first is that the sum is dominated by the final term (optimistic), and then we have an exact $T^4$ scaling, 
 \be
 I_\mathrm{tot} < \frac{N A^2}{\pi^2} \left( \frac{T_\mathrm{tot}}{N} \right)^4 \sim T_\mathrm{tot}^4.
 \ee
 A (very) pessimistic limit is that the total time is $n N T_n$, which would say each measurement takes as long as the last one.  This would give an approximate results for the last time in terms of the total time,
 \be
 T_n > \frac{T_\mathrm{tot}}{N  \log_2 \ln (T_\mathrm{tot}/N)}, 
 \ee
 which bounds the total information as
 \be
 I_\mathrm{tot} > \frac{N A^2}{\pi^2} \left( \frac{T_\mathrm{tot}}{N  \log_2 \ln (T_{tot}/N) } \right)^4 \sim (T_\mathrm{tot} /\log_2 \ln T_\mathrm{tot}) ^4,
 \ee
 which also yields $T^4$ scaling for large $T$. \\
 
 \noindent {\bf V. Rabi spectroscopy.}

 The optimal control for frequency estimation consists of $\pi$-pulses applied every half-period of the oscillating Hamiltonian and bears resemblance to a continuous drive with a Rabi frequency of $\omega$ which is the basis of Rabi spectroscopy. In this section, we compare the optimal control strategy to the classic technique of Rabi spectroscopy.   

We consider a resonant Rabi drive at the qubit transition frequency that produces Rabi oscillations at frequency $\omega_0$.  In the frame rotating with this drive, the time dependent Hamiltonian constitutes an additional Rabi drive at detuning $\omega-\omega_0$.  The transition probability for a qubit experiencing this Rabi drive for a time $T$ is given by, 
\begin{align}
P = \frac{1}{1+ (\omega-\omega_0)^2/A^2} \sin^2\left(\frac{T}{2} \sqrt{A^2+(\omega-\omega_0)^2}\right)
\end{align}
Assuming that the detuning is small, we can ignore the prefactor.  The best sensitivity is achieved if we use a detuning $(\omega-\omega_0)$ such that $P=1/2$, at the side of the first fringe.  
The quantum fisher information can be calculated from $\partial \phi/\partial \omega$, (QFI ($\propto  (d\phi/d\omega)^2$)).  The accumulated phase is $\phi = T\sqrt{A^2+(\omega-\omega_0)^2}$, therefore,
\begin{align}
\frac{\partial \phi}{\partial \omega} = \frac{T (\omega-\omega_0)}{\sqrt{A^2+(\omega-\omega_0)^2}}.
\end{align}
The optimal detuning can be determined from the frequency $\omega_0$ that yields the side of the first fringe, $T\sqrt{A^2+(\omega-\omega_0)^2} = A T+\pi/2$.   Evaluating $\partial \phi/\partial \omega$ at this optimal detuning, we have:
\begin{align}
 I_\omega^Q =\pi T^2 \frac{\pi+ 4 A T}{(\pi + 2AT)^2}.
\end{align}
We see that Rabi spectroscopy yields a worse scaling for the quantum Fisher information compared to Ramsey spectroscopy.  \\ 
%
%

\noindent{\bf VI. Statistical information and further analysis details.}

The source of error in the frequency sensitivity presented in Figure 2e is due to the estimated standard deviation of the fit parameters in determining the slope $\mathrm d\phi/\mathrm{d} \omega$ (e.g.  Fig. 2c, for which $N=10,000$ measurements are averaged for each value of $\omega$, and $101$ values of $\omega$ are incorporated into the fit). The range of $\omega$ used in determining the slope is typically $\pm 0.2\%$ and this range is increased at short interaction times to $\pm 2\%$ where the slope is small.  To measure the phase uncertainty versus number of measurements (Fig. 2d), a variable number of identical measurements ($N$) are performed to determine the mean value of $\phi$, by repeating this measurement $101$ times, the variance of this phase estimation is determined.  In Figure 3, the uncertainty resulting from the slope  is smaller than the size of the data points. 

\end{document}